\documentclass
[superscriptaddress,secnumarabic,amssymb,amsmath,nobibnotes,aps,prd,showkeys,showpacs,twocolumn,nofootinbib]{revtex4-1}%

\usepackage{bm}
\usepackage{mathrsfs}
\usepackage[all]{xy}
\usepackage{epsfig,subfigure}
\usepackage{latexsym, amssymb, amsmath, amsfonts}
\usepackage[english]{babel}
\usepackage[OT1]{fontenc}
\usepackage[utf8]{inputenc}
\usepackage{makeidx}
\usepackage[colorlinks=true,breaklinks=true]{hyperref}
\usepackage{graphicx, wrapfig, epsf}
\usepackage{float}
\usepackage{siunitx}
\usepackage[dvipsnames, table]{xcolor}
\hypersetup{urlcolor=BlueViolet,
	        citecolor=Plum,
	        linkcolor=OliveGreen}
\usepackage{fontawesome}

\definecolor{Gray}{rgb}{.9,.9,.9}

\newcommand{\aap}{{Astron.~Astrophys.}}
\newcommand{\araa}{{Annu.~Rev.~Astron.~Astrophys.}}

\newcommand{\mnras}{{MNRAS}}
\newcommand{\jcap}{{JCAP}}
\newcommand{\aj}{{Astronomic. J.}}

\newcommand{\physrep}{{Phys.~Rep.}}

\begin{document}

\title{Cooling process of brown dwarfs in Palatini $f(\mathcal R)$ gravity}

\author{Mar\'ia Benito}
\email{mariabenitocst@gmail.com}
\affiliation{National Institute of Chemical Physics and Biophysics, R\"avala 10, Tallinn 10143, Estonia
}

\author{Aneta Wojnar}
\email{aneta.magdalena.wojnar@ut.ee}
\affiliation{Laboratory of Theoretical Physics, Institute of Physics, University of Tartu,
W. Ostwaldi 1, 50411 Tartu, Estonia
}

\begin{abstract}
We present an analytical model for the evolution of brown dwarfs in quadratic Palatini $f(\mathcal R)$ gravity. We improve previous studies by adopting a more realistic description of the partially-degenerate state that characterizes 
brown dwarfs. 
Furthermore, we take into account the hydrogen metallic-molecular phase transition between the interior of the brown dwarf and its photosphere. For such improved model, we revise the cooling process of sub-stellar objects.\hspace{0.01cm} \href{https://github.com/mariabenitocst/brown_dwarfs_palatini}{\large\faGithub}
\end{abstract}

\maketitle

\section{Introduction}

Dark Matter (DM) provides a consistent explanation of gravitational phenomena at spatial scales that roughly span up to 10 orders of magnitude. It explains the Cosmic Microwave Background power spectrum \cite{2020A&A...641A...6P} and the formation of structures in the Universe \cite{2013MNRAS.432..743N}. Furthermore, DM is invoked to explain the mismatch between the observed dynamical mass and that inferred from observations of the visible component in Galaxy clusters \cite{2007MNRAS.379..209S}, elliptical and spiral galaxies \cite{2001ARA&A..39..137S}, and in dwarf and ultra-faint dwarf galaxies \cite{2015AJ....149..180O, 2019MNRAS.486.2535D}. 
Nonetheless, the nature of the DM remains unknown and none of the proposed candidates has been detected so far.

An alternative proposal to explain the observed discrepancy in the data is to modify the theory of gravity. Several modifications of gravity have been proposed in the literature (e.g. \cite{1983ApJ...270..365M, 2004PhRvD..70h3509B, 2006JCAP...03..004M, 2010RvMP...82..451S,allemandi2004accelerated,allemandi2005dark,ferraro2007modified}). These proposals are able to describe gravitational phenomena at different scales, such as the rotation curve in spiral galaxies \cite{2015PhRvD..92h4046I, 2018PhRvD..98j4061N}, the accelerating expansion of the Universe \cite{nojiri2017modified} or issues related to stellar structure (e.g \cite{olmo2011palatini,nojiri2011unified,capozziello2011extended,reva}). However, it is uncertain whether modifications of gravity are able to provide a coherent explanation at all scales. In this work, we focus on Palatini $f(\mathcal{R})$ gravity, and in particular in the Starobinsky (quadratic) model. For this theory of gravity, we present an analytical study of the evolution of brown dwarfs (BDs).

In vacuum, Palatini gravity --independently of the $f(\mathcal R)$ model-- turns out to be Einstein's theory with a cosmological constant \cite{allemandi2004accelerated,allemandi2005dark,barraco2002f}. Moreover, the
field equations are second order partially-differential equations (PDE) with respect to the metric $f(R)$ gravity (the metric formalism gives 4th order PDE with respect to the metric\footnote{However, the field equations can be rewritten as 2nd order PDE with respect to the metric, and a dynamical equation for a scalar field, which gives an extra degree of freedom (e.g. \cite{2010RvMP...82..451S,BSS}). }). There is no extra degree of freedom and, the most important feature for our purposes, is that the stellar equations are changed. Furthermore, Palatini gravity alters the early Universe physics, explains the acceleration expansion of the late Universe, provides different black hole solutions or produces wormholes with no exotic fluid \cite{Borowiec:2015qrp,szydlowski2016sewn,SSB,olmo2020junction,Lobo:2020vqh,Rubiera-Garcia:2020gcl,Olmo:2017qab,Bejarano:2016gyv,Bambi:2015zch,Olmo:2015axa,Olmo:2015xwa,Bazeia:2014poa,Lobo:2013ufa,Olmo:2012nx,Olmo:2011ja,Barragan:2009sq,Jarv:2020qqm,Wojnar:2017tmy,Wojnar:2020fqi,Wojnar:2020txr,aneta5,Wojnar:2020ckw}. 
It passes current solar system tests \cite{bonino2020solar}
since the modifications of energy and momentum appearing in Euler equation turn out not to be sensitive enough to the experiments performed for the solar system orbits \cite{junior}. The situation may however change when experiments on an atomic level will be 
available \cite{sch,ol1,ol2}.

Apart from compact stars 
\cite{TheLIGOScientific:2017first, Berti:2015itd, sun1, sun2, lina, as, craw}
and black holes, whose properties are not yet fully understood \cite{NSBH, straight}, there exists a class of stellar and sub-stellar objects which turn out to be well suited to test gravitational theories (e.g. \cite{chang}). In particular, BDs are sub-stellar objects that are not massive enough to sustain stable hydrogen burning, thus they cool down as they age. The gravitational force in BDs is balanced by electron degeneracy pressure in their cores and thermal pressure in their atmospheres. These partially-degenerate objects have masses smaller than the hydrogen-minimum mass which, accordingly to evolutionary models in GR, is $\sim \SI{0.075}{M_{\odot}}$ for a solar composition \cite{chab3}. 
BDs emit mainly in the infrared.
Since they have low luminosities, they are difficult to observe and they are mostly detected in the solar neighborhood. With the advent of wide-field surveys, large and homogeneous samples of BDs have been recently constructed \cite{2016A&A...589A..49S, 2018A&A...619L...8R, 2019ApJ...870..118S, 2019MNRAS.489.5301C, 2020arXiv201015853B}. Thus enabling statistical analyses of BDs that could constrain structural properties of our Galaxy \cite{2019MNRAS.489.5301C}, the sub-stellar mass function or sub-GeV DM particle models \cite{2020arXiv201000015L}. 

In this work, we analytically study the time evolution of BDs in quadratic Palatini $f(\mathcal{R})$ gravity. We improve previous studies by including a better description of the partially-degenerate state that characterizes BDs. The structure of the paper is as follows: in section~\ref{sec:Palatini_basics} we discuss the basic elements of Palatini $f(\mathcal{R})$ gravity and the analytical model for the time evolution of BDs is presented in section~\ref{sec:BD}. Finally, we conclude in section~\ref{sec:conclusions}. 

\section{Palatini $f(\mathcal{R})$ cheat sheet}
\label{sec:Palatini_basics}

Let us briefly recall the basic elements of Palatini gravity, which is the 
simplest example of metric-affine theories of gravity. Instead of taking the linear in $\mathcal{R}$ Lagrangian, we will consider an arbitrary, but analytical \cite{junior}, functional $f(\mathcal{R})$. The action is then written as
\begin{equation}
S=S_{\text{g}}+S_{\text{m}}=\frac{1}{2\kappa}\int \sqrt{-g}f(\mathcal{R}) d^4 x+S_{\text{m}}[g_{\mu\nu},\psi_m],\label{action}
\end{equation}
where $\mathcal{R}=\mathcal{R}^{\mu\nu}g_{\mu\nu}$ is the Ricci scalar constructed with the metric $g_{\mu\nu}$ and the Ricci tensor $\mathcal{R}^{\mu\nu}$. The latter is a function of the independent
connection $\hat\Gamma$. Adopting this, we abandon the common assumption on $g$-metricity and thus, the connection $\hat \Gamma$ might be independent of the metric $g_{\mu\nu}$. Before going further, let us comment that we use the $(-+++)$ metric signature convention and we follow the Weinberg's $\kappa=-\frac{8\pi G}{c^4}$ \cite{weinberg}.

The variation of the action (\ref{action}) with respect to the metric $g_{\mu\nu}$ provides the following field equations
\begin{equation}
f'(\mathcal{R})\mathcal{R}_{\mu\nu}-\frac{1}{2}f(\mathcal{R})g_{\mu\nu}=\kappa T_{\mu\nu},\label{structural}
\end{equation}
where $T_{\mu\nu}$ is the energy 
momentum tensor of the matter field, i.e. 
\begin{equation}
    T_{\mu\nu}=-\frac{2}{\sqrt{-g}}\frac{\delta S_m}{\delta g_{\mu\nu}}.
\end{equation}
The perfect fluid energy tensor will be assumed to describe low-mass stars.
The prime in equation~\eqref{structural} denotes derivativation with respect to the 
function's argument, that is, $f'(\mathcal{R})=\frac{df(\mathcal{R})}{d\mathcal{R}}$.

On the other hand,
varying the action with respect to the connection $\hat\Gamma$ provides
\begin{equation}
\hat{\nabla}_\beta(\sqrt{-g}f'(\mathcal{R})g^{\mu\nu})=0,\label{con}
\end{equation}
which indicates that $\hat{\nabla}_\beta$ is the covariant derivative obtained with respect to $\hat\Gamma$. Following this, we notice that there exists a conformal metric
\begin{equation}\label{met}
h_{\mu\nu}=f'(\mathcal{R})g_{\mu\nu},
\end{equation}
for which $\hat\Gamma$ is the Levi-Civita connection.
 
The trace of equation \eqref{structural} taken with respect
to $g_{\mu\nu}$ provides the structural equation
\begin{equation}
f'(\mathcal{R})\mathcal{R}-2 f(\mathcal{R})=\kappa T,\label{struc}
\end{equation}
where $T$ is the trace of the energy-momentum tensor $T_{\mu\nu}$. When a suitable functional form of $f(\mathcal R)$ is chosen, 
it is possible to solve the structural equation~\eqref{struc} in order to obtain the relation between the Palatini-Ricci curvature scalar $\mathcal{R}$ and the energy momentum trace $T$, i.e. $\mathcal{R}=\mathcal{R}(T)$. An important feature of Palatini gravity, independently of the $f(\mathcal{R})$ form, is that in vacuum --as derived from equation~\eqref{struc}-- the Einstein's vacuum solution with the cosmological constant is recovered. 

One can rewrite the field equations (\ref{structural}) as dynamical equations for the conformal
 metric $ h_{\mu\nu}$ \cite{BSS,SSB} and the undynamic scalar 
 field
 denoted as $\Phi=f'(\mathcal{R})$:
 \begin{subequations}
	\begin{align}
	\label{EOM_P1}
	 \bar R_{\mu\nu} - \frac{1}{2} h_{\mu\nu} \bar R  &  =\kappa \bar T_{\mu\nu}-\frac{1}{2} h_{\mu\nu} \bar U(\Phi)
	\end{align}
	\begin{align}
	\label{EOM_scalar_field_P1}
	  \Phi\bar R &  -  (\Phi^2\,\bar U(\Phi))^\prime =0
	\end{align}
\end{subequations}
where $\bar U(\Phi)=\frac{\mathcal{R}\Phi-f(\mathcal{R})}{\Phi^2}$ and the energy momentum 
tensorin the Einstein's frame is $\bar T_{\mu\nu}=\Phi^{-1}T_{\mu\nu}$. It was demonstrated in several works \cite{Wojnar:2017tmy,aneta,o,o1,o2} that this representation of 
the Palatini $f(\mathcal R)$ gravity simplifies examinations of physical problems.

In our work we will focus on the quadratic (Starobinsky) functional form of $f(\mathcal{R})$, i.e.
\begin{equation}
    f(\mathcal R)=\mathcal{R}+\beta\mathcal{R}^2,
\end{equation}
where $\beta$ is a the Starobinsky parameter with dimension $[\text{m}^{-2}]$. Later on, we will introduce the parameter $\alpha$ which is related to the Starobinsky parameter $\beta$.

\section{Brown dwarf's model}
\label{sec:BD}
In this section we improve the brown dwarf's analytical 
model for Palatini $f(\mathcal R)$ gravity considered in \cite{gonzalo}. The main difference with respect to the previous work is related to the Equation of State (EoS). The model discussed in \cite{gonzalo} used the polytropic EoS which works well in the degenerate and ideal gas extremes, being however a poor description for the intermediate zone, when one deals with a mixture of these two gases \cite{burr}. In this work we consider the EoS first presented in \cite{auddy} which better describes a mixture of degenerate and ideal gas states at finite temperature.
We then provide a simple cooling model for these sub-stellar objects.

\subsection{Equation of State for a partially-degenerate Fermi gas}

The barotropic EoS $p=p(\rho)$ --where $p$ and $\rho$ are the pressure and energy density, respectively-- which accounts for a mixture of a degenerate Fermi gas of electrons at a finite temperature and a gas of ionized hydrogen and helium is given as follows \cite{auddy}
\begin{align}\label{fermi}
    p&=C \left( \frac{\rho}{\mu_e} \right)^\frac{5}{3} \left[1-\frac{5}{16}\Psi \rm{ln}(1+e^{-1/\Psi})  \right.\nonumber \\
    &\left. +  \frac{15}{8}\Psi^2\left( \frac{\pi^2}{3}+\rm{Li}_2[-e^{-1  /\Psi}]\right)+a\Psi\right],
\end{align}
with $C=10^{13} \rm{cm}^4g^{-2/3}s^{-2}$, $\rm{Li}_2$ denotes the second order polylogarithm function and the number of baryons per electron is given by $1/\mu_e=X+Y/2$, where $X$ and $Y$ are the mass fractions of hydrogen and helium, respectively.
The degeneracy parameter $\Psi$ is defined as 
\begin{equation}\label{deg}
    \Psi=\frac{k_B T}{\mu_F}=\frac{2m_ek_B T}{(3\pi^2\hbar^3)^{2/3}}\left( \frac{\mu_e}{\rho_c N_A} \right)^{2/3},
\end{equation}
where $\mu_F$ is the electron's Fermi energy in the degenerate limit, $T$ is the gas temperature, $\rho_c$ is the density of the BDs' core and the rest constants have the usual meaning. Finally, the quantity $a=\frac{5}{2}\mu_e\mu_1^{-1}$ with $\mu_1$ defined as
\begin{equation}
\frac{1}{\mu_1}=(1+x_{H^+})X+\frac{Y}{4},
\end{equation}
where $x_{H^+}$ is the ionization fraction of hydrogen. This fraction changes from the completely ionized core to the surface of the BD, which is composed of molecular hydrogen and helium \cite{auddy}. Its values depends on the phase transition points \cite{chab4} to which we will come back later.

The EoS~\eqref{fermi} has a familiar polytropic form for $n=3/2$
\begin{equation}\label{pol}
    p=K\rho^{1+\frac{1}{n}}
\end{equation}
with $K=C\mu_e^{-\frac{5}{3}}(1+b+a \Psi)$ and 
\begin{equation}\label{defb}
    b=-\frac{5}{16}\Psi \rm{ln}(1+e^{-1/\Psi})+\frac{15}{8}\Psi^2\left( \frac{\pi^2}{3}+\rm{Li}_2[-e^{-1  /\Psi}] \right),
\end{equation}
which takes into account the corrections due to the finite temperature of the gas.
Since we are interested in BDs, the polytropic models with $n=3/2$ together with such improvements are well-suited to describe these sub-stellar objects. 

The hydrostatic equilibrium equations in the Einstein frame was shown to be in our case \cite{aneta}
\begin{equation}\label{part_EL}
 -\bar{r}^2\Phi(\bar{r})\frac{\rm d}{\rm{d}\bar{r}}p=G\mathcal{M}(\bar{r})\rho(\bar{r}),\;\;\mathcal{M}(\bar{r})\approx\int^{\bar{r}}_0 4\pi\rho \tilde{r}^2d\tilde{r},
\end{equation}
which can be further rewritten,
after introducing the standard dimensionless variables
\begin{align}
 \bar{r}&=r_c\bar{\xi},\;\;\;\rho=\rho_c\theta^n,\;\;\;p=p_c\theta^{n+1},\label{def1}\\
 r^2_c&=\frac{(n+1)p_c}{4\pi G\rho^2_c},\label{def2}
\end{align}
as the Lane-Emden equation, which for quadratic Palatini $f(\mathcal R)$ gravity after coming back to Jordan frame ($\bar{\xi}^2=\Phi \xi^2$) is given by:
\begin{equation}\label{LE}
 \frac{1}{\xi}\frac{d^2}{d\xi^2}\left[\sqrt{\Phi}\xi\left(\theta-\frac{4\kappa^2 c^2\rho_c\alpha}{5}\theta^\frac{5}{2}\right)\right]=
 -\frac{(\Phi+\frac{1}{2}\xi\frac{d\Phi}{d\xi})^2}{\sqrt{\Phi}}\theta^\frac{3}{2},
\end{equation}
where $\Phi=1+2\alpha \theta^\frac{3}{2}$, $\alpha=\kappa c^2\beta\rho_c$ and $\kappa=-\frac{8\pi G}{c^4}$ \cite{weinberg}. For simplification, from now on we will use the parameter $\alpha$ instead of the Starobinsky parameter $\beta$. Moreover, the range of $\alpha$ is $(-0.5;+\infty)$. The parameters $\rho_c$ and $p_c$ stands for central density and central pressure, respectively. In such framework, 
the temperature can be expressed as $T=T_c\theta(\xi)$, with $T_c$ being the central temperature, while the density is  
$\rho=\rho_c\theta^{3/2}(\xi)$. 
The function $\theta(\xi)$ is the solution of the (modified) Lane-Emden equation with respect to the radial coordinate 
$\xi=r\rho_c\sqrt{8\pi G/(2p_c)}$. The solution $\theta(\xi)$ crosses zero at $\xi_R$ which corresponds to the dimensionless BD's radius. For a more detailed discussion see e.g. \cite{aneta}.

The BD's radius $R$, central density $\rho_c$, and pressure $p$ can be obtained by numerically solving equation~\eqref{LE}. For an arbitrary polytropic parameter $n$, these parameters are expressed as 
\begin{align}
 R&=\gamma_n\left(\frac{K}{G}\right)^\frac{n}{3-n}M^\frac{n-1}{n-3} \label{radiuss},\\
 \rho_c&=\delta_n\left(\frac{3M}{4\pi R^3}\right) \label{rho0s} ,\\
 p&=K\rho_c^{\frac{n+1}{n}}\theta^{n+1}.\label{presc}
\end{align}
The values of the parameters $\gamma_n$ and $\delta_n$ depend on the adopted theory of gravity. In the case of Palatini $f(\mathcal{R})$ gravity, they take the following forms \cite{artur}
\begin{align}
  \gamma_n&=(4\pi)^\frac{1}{n-3}(n+1)^\frac{n}{3-n}\omega_n^\frac{n-1}{3-n}\xi_R,\label{gamma}\\
 \delta_n&=-\frac{\xi_R}{3\frac{\Phi^{-\frac{1}{2}}}{1+\frac{1}{2}\xi\frac{\Phi_\xi}{\Phi}}\frac{d\theta}{d\xi}\mid_{\xi=\xi_R}}, \label{delta} \\ 
  \omega_n&=-\frac{\xi^2\Phi^\frac{3}{2}}{1+\frac{1}{2}\xi\frac{\Phi_\xi}{\Phi}}\frac{d\theta}{d\xi}\mid_{\xi=\xi_R}.\label{omega}
\end{align}
Using the above definitions, the stars' external and internal characteristics can be obtained for
the EoS given by equation~\eqref{fermi} as functions of the mass $M$ and the degeneracy parameter $\Psi$. That is,\footnote{We drop the sub-index $3/2$ from $\gamma_{3/2}$ and $\delta_{3/2}$ for convenience.}
\begin{align}
    R=&1.19141\times 10^9\gamma\left(\frac{M_\odot}{M}\right)^{\frac{1}{3}}\mu_e^{-\frac{5}{3}}(1+b+a\Psi)\,\,[\rm{cm}],\label{radius}\\
    \rho_c=&2.80791\times 10^{5}\frac{\delta}{\gamma^3}\left(\frac{M}{M_\odot}\right)^2\frac{\mu_e^5}{(1+b+a\Psi)^3}\,\,[g/\rm{cm}^3],\label{ro}\\
    p_c=&1.20403\times 10^9\frac{\delta^{5/3}}{\gamma^5}\left(\frac{M}{M_\odot}\right)^{10/3}\frac{\mu_e^{20/3}}{(1+b+a\Psi)^4}\,\,[\rm{Mbar}].
\end{align}
Furthermore, the central temperature takes the following form:
\begin{equation}\label{tem}
    T_c=1.29396\times10^9\frac{\delta^{2/3}}{\gamma^2}\left(\frac{M}{M_\odot}\right)^{4/3}\frac{\mu_e^{8/3}}{\Psi(1+b+a\Psi)^2}\,\,[\rm{K}]
\end{equation}
when one uses the equation (\ref{ro}) together with the definition of the degeneracy parameter (\ref{deg}). Let us just emphasise that in the above formulas, the values of $\gamma$ and $\delta$ depend on the solution of the modified Lane-Emden equation \eqref{LE} with respect to the value of $\alpha$.

\subsection{Brown dwarfs' surface properties}\label{ssurf}
Although BDs are simpler than compact objects, there are still missing elements in the theoretical and numerical modelling of the BD's interior. This introduces sizeable uncertainties in the predicted surface temperature. Despite this, there exist models which allow to express the surface temperature in an analytical form -- that is based on the isentropic BD's interior and the phase transition between the interior and the photosphere-- which is convenient for our purposes. In \cite{chab4} it was shown that a first order phase transition for the metallization of hydrogen happens for pressure and temperatures suitable for giant planets and BDs. 
The effective temperature $T_\text{eff}$ can be written then in terms of the degeneracy parameter $\Psi$ and the photospheric density $\rho_\text{ph}$ as \cite{auddy}
\begin{equation}\label{tsur}
    T_\text{eff}=b_1 \times 10^6 \rho_\text{ph}^{0.4}\Psi^\nu\,\,\text{K},
\end{equation}
where the values of the parameters $b_1$ and $\nu$ depend on the specific model adopted for describing the phase transition between a metallic hydrogen and helium state that characterizes the BD's interior and the photosphere, which is composed of molecular hydrogen and helium. We adopt different models presented in \cite{chab4} which are summarized in table \ref{tab0}.
\bgroup
\def\arraystretch{1.3}
\begin{table}[h]
\centering
\begin{tabular}{|c||c |c|c|}
\hline
Model &  $x_{H^+}$  & $b_1$ & $\nu$ \\
\hline\hline
A  & 0.240 & 2.87 &  1.58 \\
B  & 0.250 & 2.70  & 1.59 \\
C  &  0.250 & 2.26 & 1.59 \\
D   &  0.255  & 2.00  & 1.60 \\
E  &  0.260 & 1.68  & 1.61 \\
F  &  0.250 & 1.29  & 1.59 \\
G  &  0.165 & 0.60  & 1.44 \\
H  &  0.090 & 0.40  & 1.30 \\
\hline
\end{tabular}
\caption{Different metallic-molecular phase transition points taken from \cite{chab4}.}
\label{tab0}
\end{table}

The surface temperature, given by equation~\eqref{tsur}, is obtained from matching the entropy in the BD's interior,
\begin{equation}\label{entr}
 S_\text{interior}=\frac{3}{2}\frac{k_BN_A}{\mu_{1mod}}(\rm{ln}\Psi+12.7065)+C_1,
\end{equation}
where $C_1$ is an integration constant of the first law of thermodynamics while 
\begin{equation}
 \frac{1}{\mu_{1mod}}=\frac{1}{\mu_1}+\frac{3}{2}\frac{x_{H^+}(1-x_{H^+})}{2-x_{H^+}},
\end{equation}
with the photospheric entropy of non-ionized molecular hydrogen and helium mixture \cite{auddy}. 
The detailed derivation of this temperature can be found in \cite{auddy} and 
\cite{Burrows:1992fg,stev}.

In order to estimate the surface luminosity, we will follow the approach presented 
in \cite{Burrows:1992fg}. The surface of a star can be assumed to lie at the photosphere which is  
defined at the radius for which the optical depth 
equals $2/3$, i.e.
\begin{equation} \label{eq:od}
 \tau(r)=\kappa_R\int_r^\infty \rho dr=\frac{2}{3},
\end{equation}
where $\kappa_R$(=$\SI{0.01}{cm^2/g}$) is Rosseland's mean opacity. Since the radius of the photosphere is very close to the stellar radius, we approximate the surface gravity as a constant. That is,
\begin{equation}\label{surf}
 g\equiv\frac{G m(r)}{r^2}\sim\frac{GM}{R^2}=\text{const},
\end{equation}
where $M=m(R)$.
Then, the hydrostatic equilibrium for the Palatini quadratic model is written as \cite{gonzalo}
\begin{equation}\label{pres}
 p'=-g\rho(1+\kappa c^2 \beta [r\rho'-3\rho]),
\end{equation}
where now $'\equiv d/dr$ in Jordan frame.  
The mass function of a non-relativistic star in our model can be approximated to the 
familiar form $ m'(r)=4\pi r^2\rho(r)$ allowing to write
 \begin{equation}
  m''=8\pi r\rho+4\pi r^2 \rho',
\end{equation}
where the second derivation of the mass $m$ is given by differentiating equation~\eqref{surf}. Using this in equation~\eqref{pres} we may write,
\begin{equation}\label{hyd}
 p'=-g\rho\left( 1+8\beta\frac{g}{c^2 r} \right),
\end{equation}
which with the help of equation~\eqref{eq:od} we integrate to the following form
\begin{equation} \label{presPH}
 p_{ph}=\frac{2}{3\kappa_R}\frac{GM\left( 1+8\beta\frac{GM}{c^2 R^3} \right)}{R^2}.
\end{equation}

The EoS \eqref{fermi} near to the photosphere, where the degeneracy is negligible, provides the photospheric pressure in the ideal gas form
\begin{equation}\label{ideal}
    p_\text{ph}=\frac{\rho_\text{ph}N_A k_B T_\text{eff}}{\mu_2}
\end{equation}
with $1/\mu_2=X/2+Y/4$.
Using the radius given by the equation~\eqref{radius} and the central density in equation~\eqref{ro}, the photospheric pressure can be written as a function of mass $M$ and the degeneracy parameter $\Psi$ as
\begin{equation}
    p_\text{ph}=\frac{62.3488}{\kappa_R
    \gamma^2}
    \left(\frac{M}{M_\odot} \right)^{5/3}\frac{\mu_e^{10/3}(1-1.33\frac{\alpha}{\delta})}{(1+b+a\Psi)^2}
    \,\text{bar}.
\end{equation}
Combinin this with the equation \eqref{ideal} and the surface temperature given by \eqref{tsur}, we may express, in a similar way, the photospheric density as a function of the mass and the degeneracy parameter, i.e.
\begin{equation}
   \rho_\text{ph}= \frac{62.3488}{\kappa_R\gamma^2 N_A k_B}\left(\frac{M}{M_\odot} \right)^{5/3}\frac{\mu_e^{10/3}\mu_2(1-1.33\frac{\alpha}{\delta})}{(1+b+a\Psi)^2b_1\Psi^\nu}
    \,\frac{\text{g}}{\text{cm}^3}.
\end{equation}
Finally, the photospheric temperature has the following form:
\begin{align}
    T_\text{eff}&=\frac{2.558\times10^4\,\text{K}}{\kappa_R^{0.286}}
    \left(\frac{M}{M_\odot} \right)^{0.4764}
     \frac{\Psi^{0.714\nu}}{(1+b+a\Psi)^{0.571}}\nonumber\\
    &\times
    b_1^{0.714}\left(1-1.33\frac{\alpha}{\delta}\right)^{0.286}\gamma^{-0.572},
\end{align}
where $\mu_e=1.143$ and $\mu_2=2.286$ were used.

Assuming black body radiation and using the Stefan-Boltzman law $L=4\pi R^2\sigma T^4_\text{eff}$, where $\sigma$ is the Stefan-Boltzmann constant, we easily obtain the BDs' luminosity as a function of mass and the degeneracy parameter in the following form
\begin{align}\label{lumph}
    L&=\frac{0.0721 L_\odot}{\kappa_R^{1.1424}}
    \left(\frac{M}{M_\odot} \right)^{1.239}
    \frac{\Psi^{2.856\nu}}{(1+b+a\Psi)^{0.2848}}\nonumber\\
    &\times
    b_1^{2.856}\left(1-1.33\frac{\alpha}{\delta}\right)^{1.143}\gamma^{-0.286}.
\end{align}

\subsection{Cooling model for brown dwarfs}
In order to express the luminosity  (\ref{lumph}) as a function of time $t$, we need to find out an evolutionary equation for the the degeneracy parameter $\Psi$. Following the steps of \cite{burr} and \cite{stev}, together with the improved EoS first obtained in \cite{auddy} and used in this work, the evolution of the luminosity of BDs as a function of time for Palatini gravity can be found. 

Applying the energy equation from the first and the second law of thermodynamics we may describe the pace of cooling and contraction related to such objects as
\begin{equation}
    \frac{\rm d E}{\rm d t}+p\frac{\rm d V}{\rm d t}=T\frac{\rm d S}{\rm d t}
    =\dot\epsilon-\frac{\partial L}{\partial M}
\end{equation}
where $S$ is the entropy per unit mass while other symbols have standard meaning. BDs are not massive enough to sustain stable hydrogen burning, therefore, they cool as they age and the energy generation term $\dot\epsilon$ can be ignored. Integrating the previous equation over mass one has
\begin{equation}
    \frac{\rm d\sigma}{\rm dt} \left[
    \int N_A k_B T \rm dM
    \right]=-L,
\end{equation}
where $L$ is a surface luminosity and $\sigma=S/k_BN_A$. Using equation~\eqref{deg} and the polytropic relation given by~\eqref{pol} in order to get rid of $T$ and $\rho$, we may write 
\begin{equation}
    \frac{\rm d\sigma}{\rm dt}
    \frac{N_A A \mu_e\Psi}{C(1+b+a\Psi)} \int p \rm dV
   =-L,
\end{equation}
where $A=(3\pi\hbar^3 N_A)^\frac{2}{3}/(2m_e)\approx4.166\times10^{-11}$
while the integral in the Jordan frame is given by 
\begin{equation}
    \int p \rm dV=\frac{2}{7}\Omega G\frac{M^2}{R}
\end{equation}
with $\Omega=(\Phi^{3/2}/(1+\frac{1}{2}\xi\Phi'/\Phi))^{-4/3}$. It can be shown that $\Omega=1$ for $n=3/2$ \cite{artur,aneta5}.

From the entropy formula (\ref{entr}), the entropy rate is given simply by
\begin{equation}
     \frac{\rm d\sigma}{\rm dt}=\frac{1.5}{\mu_{1\text{mod}}}\frac{1}{\Psi} \frac{\rm d\Psi}{\rm dt}
\end{equation}
and together with the luminosity (\ref{lumph})
we may finally write down the evolutionary equation for the degeneracy parameter $\Psi$
\begin{align}
    \frac{\rm d\Psi}{\rm dt}&=-
    \frac{1.1634\times10^{-18}b_1^{2.856}\mu_{1\text{mod}}}{\kappa_R^{1.1424}\mu_e^{8/3}}
    \left(\frac{M_\odot}{M} \right)^{1.094}
    \\
    &\times\Psi^{2.856\nu}(1+b+a\Psi)^{1.715} \frac{\gamma^{0.7143}}{\Omega} \left(1-1.33\frac{\alpha}{\delta}\right)^{1.143}.\nonumber
\end{align}
We have numerically solved the above ordinary differential equation by assuming that $\Psi=1$ at $t=0$ for the parameters' values given in the Table \ref{tab}. The code for solving this differential uses the GNU Scientific Library~\cite{galassi2007gnu} and can be found at \href{https://github.com/mariabenitocst/brown_dwarfs_palatini}{this github repository}.

As it turns out, for the examined range of the parameter $\alpha$ --which are summarized in table~\ref{tab}--, the evolution of the degeneracy parameter does not differ significantly from GR ($\alpha=0$), as it can be seen from figure~\ref{fig:degeneracy_vs_time}. By plugging in the numerically obtained $\Psi(t)$ into equation~\eqref{lumph} we obtain the evolution of the luminosity of BDs as a function of time. This can be seen in figure~\ref{fig:luminosity_vs_time}.

\begin{figure*}[h!]
\includegraphics[scale=0.5]{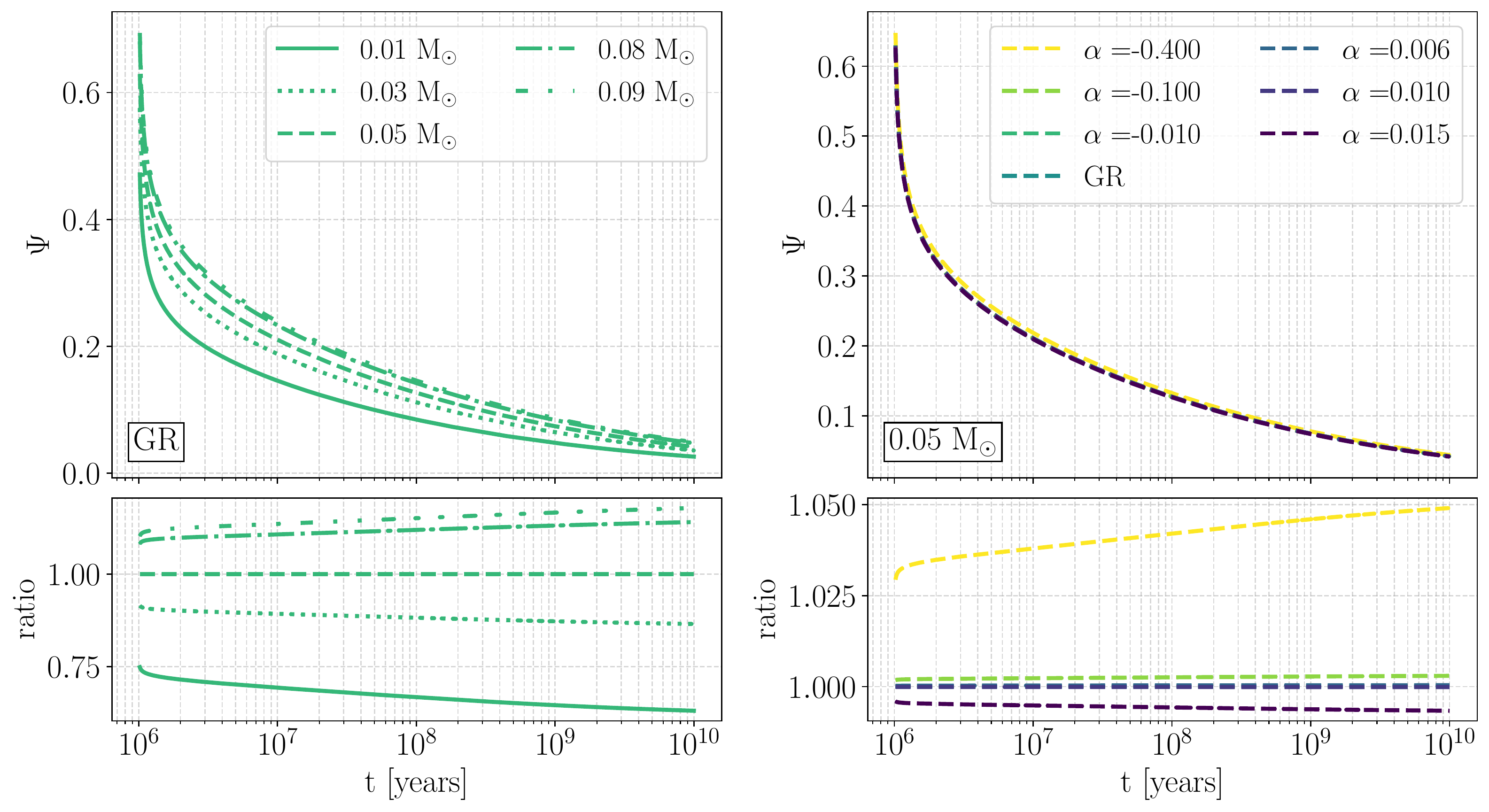} 
\caption{The time evolution of the degeneracy parameter $\Psi$ in the case of GR for different BDs masses (left panels) and for M=$\SI{0.05}{M_\odot}$ and different values of the $\alpha$ parameter (right panels). In the bottom panels, we show the ratio with respect to the curve with M=$\SI{0.05}{M_\odot}$ in the left. In the right bottom panel, the ratio of the time evolution in different Palatini models versus GR is depicted.}
\label{fig:degeneracy_vs_time}
\end{figure*}

\begin{figure*}[h!]
\includegraphics[scale=0.5]{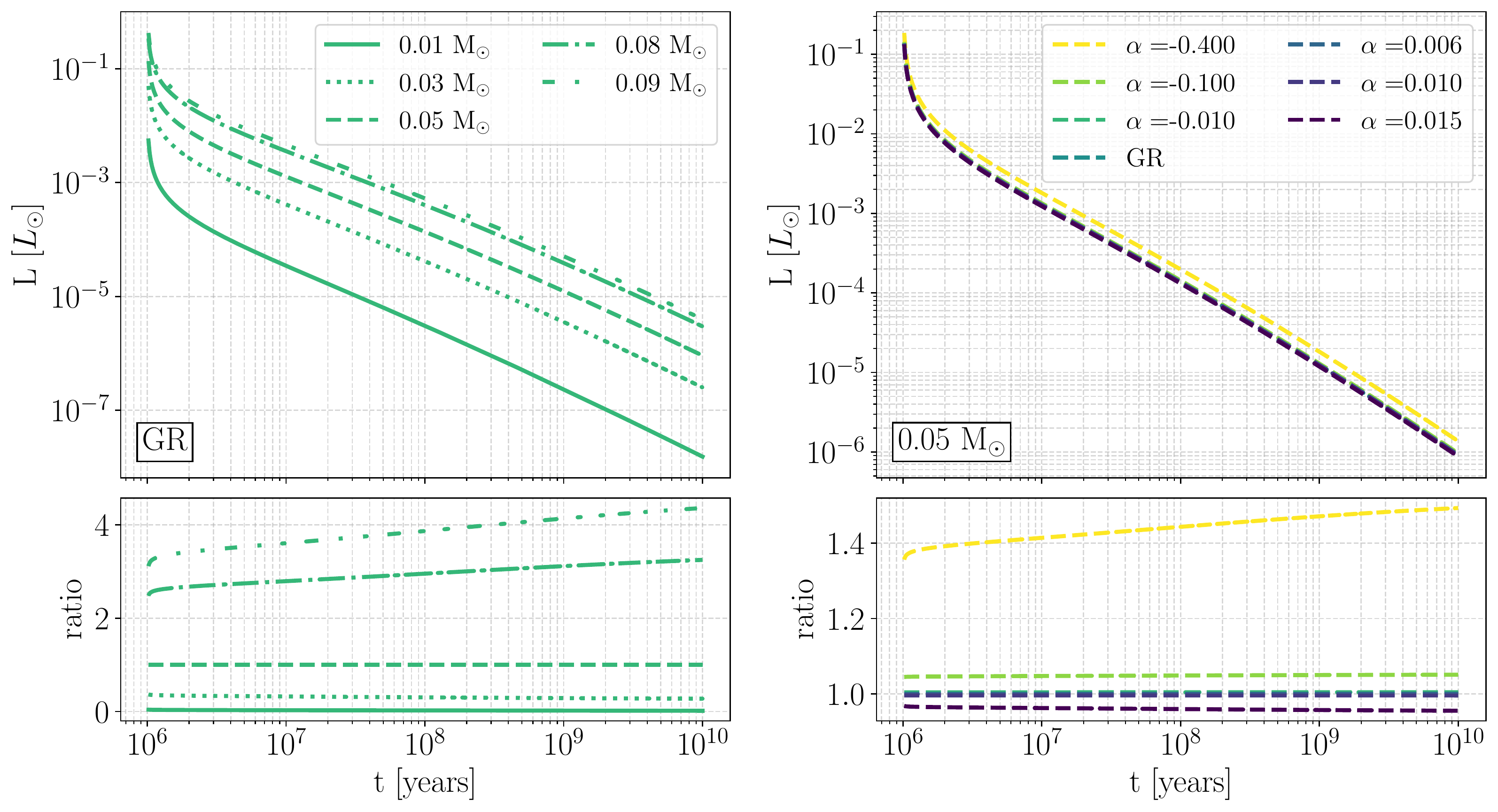} 
\caption{Time evolution of the BDs' luminosity. Left panels: assuming GR and different masses. In the bottom panel, we show the ratio with respect to our fiducal mass M=$\SI{0.05}{M_\odot}$. Right panels: for fixed M=$\SI{0.05}{M_\odot}$ and different values of the parameter $\alpha$. In the bottom panel we show the ratio with respect to the GR case.}
\label{fig:luminosity_vs_time}
\end{figure*}

\bgroup
\def\arraystretch{1.5}
\begin{table}[t!]
\begin{center}
\begin{tabular}{|c||c c|c c|}
\hline
$\alpha$ & $\xi_R$ & $\omega_{3/2}(\xi_R)$ & $\gamma_{3/2}(\xi_R)$  & $\delta_{3/2}(\xi_R)$  \\
\hline\hline
-0.400 & 3.16 &  1.40 & 1.63 & 7.53  \\
-0.100 & 3.64 & 2.39  & 2.25 & 6.67  \\
-0.010 & 3.65 & 2.68  & 2.35 & 6.09  \\
\hline
0 (GR)    & 3.65 & 2.71 &  2.36 & 5.97 \\
\hline
0.006   & 3.66 & 2.73 & 2.36  & 5.95  \\
0.010   & 3.66 & 2.75 & 2.37 & 5.93  \\
0.015   & 3.66 & 2.77 &  2.46 & 5.89  \\
\hline
\end{tabular}
\caption{Numerical values of $\xi_R$ obtained from $\theta(\xi_R)=0$, and the associated values of the functions $\gamma_{3/2}$, $\omega_{3/2}$, and $\delta_{3/2}$  for different values of $\alpha=\kappa c^2 \beta \rho_c$.}
\label{tab}
\end{center}
\end{table}

\section{Conclusions}
\label{sec:conclusions}
In this work we have updated the analytical model of BDs considered in \cite{gonzalo} by adopting the EoS first presented in \cite{auddy}. In addition, we have provided a cooling model for sub-stellar objects in quadratic Palatini gravity. This more realistic EoS describes the BDs' interior as a mixture of a degenerate Fermi gas and ions of hydrogen and helium. Our model further includes a proper treatment of the hydrogen's phase transition from the photosphere, in which the hydrogen is in its molecular form, and the interior of BDs, where the hydrogen is ionized. 
For this improved description of these partially-degenerate sub-stellar objects, we conclude that:
\begin{itemize}
\item The time evolution of the degeneracy parameter in quadratic Palatini gravity --for the values of $\alpha$ adopted in this work-- differs with respect to that in GR by $\lesssim 2.5\%$ when the BD is about 1 Myr and by $\lesssim 5\%$ at 10 Gyr.
\item The difference in the estimated BDs' luminosity between Palatini gravity and GR slightly increases with the age of the BD. Furthermore, BDs have a lower luminosity in Palatini gravity with positive $\alpha$ values ($\beta <0$) than in GR. On the other hand, for negative $\alpha$-values ($\beta >0$), gravity inside BDs is weaker and the luminosity of BDs in Palatini is larger with respect to that estimated in GR.
\item For $\alpha$-values smaller than 0.1 in absolute value, the difference between the luminosity predicted in GR and the one in Palatini gravity is smaller than 6\%. For $\alpha=-0.4$, this difference increases up to 50\%. Although this difference is significant, it is smaller than the estimated difference obtained by varying the BD's mass. For instance, a variation in mass of 40\%, produces a change in luminosity larger than a factor of 2.
\end{itemize}
To sum up, BDs could constrain modifications of gravity and, in particular, the observed luminosity of BDs might be used to constrain the $\beta$ Starobinsky parameter, as shown in this work. Current and future wide-field survey will provide large and homogeneous samples of BDs that could be used in this respect.
Our study is a first analytical step in this direction and we leave to future work the comparison of our results with more complete numerical models.

\vspace{5mm}
\noindent \textbf{Acknowledgement.} 
The authors would like to thank Rain Kipper for his comments. This work was supported by the EU through the European Regional Development Fund CoE program TK133 ``The Dark Side of the Universe." M.B. is supported by the Estonian Research Council PRG803 grant.

\bibliography{main.bib}

\end{document}